\documentclass[preprint, amsmath, amssymb, aps, prb, superscriptaddress, floatfix]{revtex4-2}
\usepackage{color}% Color fonts
\usepackage{graphicx}% Include figure files
\usepackage{dcolumn}% Align table columns on decimal point
\usepackage{bm}% bold math
\usepackage[normalem]{ulem} % for strikethrough 
\usepackage{natbib} %%

\usepackage[T1]{fontenc}
\usepackage{amsmath,amssymb}
\usepackage{siunitx}

\usepackage{dcolumn}% Align table columns on decimal point
\usepackage{bm}% bold math
\usepackage{color, colortbl}% Color fonts

\usepackage{textcomp, gensymb}
\usepackage{braket}

\DeclareSIUnit{\belmilliwatt}{Bm}
\DeclareSIUnit{\dBm}{\deci\belmilliwatt}
\DeclareSIUnit{\torr}{Torr}
\DeclareSIUnit\angstrom{\text {Å}}

\begin{document}
\title{Controlled growth of rare-earth-doped TiO$_2$ thin films on III-V semiconductors for hybrid quantum photonic interfaces}

\author{Henry C. Hammer}
\affiliation{Department of Physics \& Astronomy, The University of Iowa, Iowa City, IA 52242}

\author{Caleb Whittier}
\affiliation{Department of Materials Science \& Engineering, McMaster University, Hamilton, Ontario L8S 4L7, Canada}

\author{Nathan A. Helvy}
\affiliation{Department of Physics \& Astronomy, The University of Iowa, Iowa City, IA 52242}

\author{Christopher Rouleau}
\affiliation
{Center for Nanophase Materials Sciences, Oak Ridge National Laboratory, Oak Ridge, Tennessee 37831, USA}

\author{Nabil D. Bassim}
\affiliation{Department of Materials Science \& Engineering, McMaster University, Hamilton, Ontario L8S 4L7, Canada}
\affiliation
{Canadian Centre for Electron Microscopy, McMaster University, Hamilton, Ontario L8S 4M1, Canada}
\author{Ravitej Uppu}
\email{ravitej-uppu@uiowa.edu}
\affiliation{Department of Physics \& Astronomy, The University of Iowa, Iowa City, IA 52242}

\begin{abstract}
Quantum photonic networks require two distinct functionalities: bright single-photon sources and long-lived quantum memories. 
III-V semiconductor quantum dots excel as deterministic and coherent photon emitters, while rare-earth ions such as erbium (Er$^{3+}$) in crystalline oxides offer exceptional spin and optical coherence at telecom wavelengths. 
Combining these systems and their functionalities through direct epitaxy is challenging because of lattice mismatch and incompatible growth conditions. 
Here we demonstrate low-temperature pulsed laser deposition of Er$^{3+}$-doped TiO$_2$ thin films directly on GaAs and GaSb substrates. 
Controlled surface preparation with an arsenic cap and an oxygen-deficient buffer layer enables the growth of epitaxial anatase TiO$_2$ (001) at $\sim390~\si{\celsius}$ with sub-$300~\si{\pico\meter}$ surface roughness, while avoiding interface degradation.
In contrast, high-temperature oxide desorption or growth temperatures drive the transition to rough, polycrystalline rutile film, as confirmed by transmission electron microscopy. 
Minimal coincident interface area (MCIA) modeling explains the orientation-selective growth on GaAs and GaSb. 
Raman and cryogenic photoluminescence excitation spectroscopy verify the crystal phase and optical activation of Er$^{3+}$ ions. 
This multi-parameter growth strategy helps preserve III-V quantum dot functionality and yields smooth surfaces suitable for low-loss nanophotonic structures. 
Our results establish a materials platform for monolithically integrating rare-earth quantum memories with semiconductor photon sources, paving the way toward scalable hybrid quantum photonic chips.
\end{abstract}

\maketitle

%%%%%%%%%%%%%%%%% Introduction %%%%%%%%%%%%%%%%%%%%%

\section{Introduction}
Quantum photonic networks require material platforms that support both bright, deterministic photon sources \cite{Uppu2021, Yu2023a} and long-lived quantum memories \cite{Atature2018, Awschalom2018, Lvovsky2009}, ideally operating at telecommunication wavelengths \cite{Wehner2018,Azuma2023}.
These two functionalities impose fundamentally different material requirements \cite{deLeon2021, Neuwirth2021}.
While epitaxial III-V quantum dots (QDs) offer bright and coherent photon emission \cite{Uppu2020, Tomm2021}, rare-earth ions (REIs) in crystalline oxide hosts provide exceptionally long (millisecond-scale) spin and optical coherence times, making them well suited as quantum memories \cite{Thiel2011, Zhong2019, Goldner2025}.
Combining these complementary systems into a monolithic architecture while preserving coherence remains challenging because of their mismatched lattice structures, thermal processing, and growth chemistries. 

Heterogeneous integration methods, such as flip-chip bonding, have demonstrated functional integration of III-V or silicon nanophotonics (e.g., waveguides and cavities) to REI-doped crystals \cite{Wu2023, Dibos2018, Chen2020, Raha2020, Ourari2023, Yu2023, Uysal2025}.
However, these methods suffer from alignment complexity, uncontrolled interfaces, and bonding-induced losses that degrade coherence \cite{Wu2023}.
Direct growth of erbium-doped oxide thin films on silicon, including TiO$_2$ \cite{Dibos2022, Ji2024, Singh2024, pettit2025}, CeO$_{2}$ \cite{Grant2024, Zhang2024}, and Y$_{2}$O$_{3}$ \cite{Scarafagio2019, Singh2020, gupta2023}, have shown narrow inhomogeneous optical linewidths, including over $400~\si{\micro\second}$ spin coherence time ($T_{2}$) for scalable quantum memories.
Yet, because silicon lacks efficient single-photon sources, these platforms require extreme Purcell enhancement of the REI's optical lifetimes to approach the brightness of III-V QDs.

A more versatile strategy is to integrate REI-doped oxides directly with III-V semiconductors.
Among potential oxide hosts, TiO$_2$ is especially promising for Er$^{3+}$-based quantum memories \cite{Ferrenti2020, Kanai2022}. 
It combines a wide bandgap and high refractive index at telecom wavelengths with a nearly nuclear-spin-free lattice environment, while supporting dry-etching processes compatible with III-V nanofabrication \cite{Rams1997, Kischkat2012, Norasetthekul2001, Midolo2015}.
Er$^{3+}$ ions substitute at the Ti$^{4+}$ sites of non-polar $D_{2h}$/$D_{2d}$ symmetry in rutile/anatase phases, which suppresses permanent dipole formation and supports long optical lifetimes and narrow linewidths, recently measured in both bulk and thin-film TiO$_2$ on silicon \cite{Phenicie2019, Stevenson2022, Martins2024, Singh2024}.
Despite its promise, direct growth of Er$^{3+}$:TiO$_2$ on III-V semiconductors for quantum photonics has remained largely unexplored. 
Previous pulsed laser deposition (PLD) studies of TiO$_2$ thin films, undoped \cite{Liu2001_1, Liu2001_2} or indium-doped \cite{AlMashary2020}, on GaAs have demonstrated polycrystalline rutile (R-TiO$_2$) thin films with limited interface control.
However, these samples were not doped with REIs, and the studies lacked a systematic analysis of growth phase selectivity and interface control, both of which are critical for preserving the quantum coherence of optically activated REIs. 
Moreover, the growth temperature exceeded $500\si{\celsius}$, which is detrimental for QD functionality.

In this study, we demonstrate low-temperature heteroepitaxial growth of Er$^{3+}$:TiO$_2$ thin films on GaAs and GaSb substrates using PLD, with a focus on phase selectivity, interface control, and optical activity of erbium ions. 
We introduce interface preparation steps that facilitate low-temperature (below $400~\si{\celsius}$) growth of crystalline thin films with $<300~\si{\pico\meter}$ roughness, compatible with low-loss nanophotonic structures for coherent spin-photon interactions.
The crystalline phase of the thin films can be tuned between the anatase (A-TiO$_2$) and rutile phases (R-TiO$_2$) by controlling either the growth temperature or adapting the interface preparation steps.
Minimal coincident interface area (MCIA) analysis explains the orientation-selective growth of anatase (001) on GaAs.
Raman spectroscopy, cryogenic photoluminescence excitation (PLE), and electron microscopy confirm the crystal phase and Er$^{3+}$ optical activity. 
Together, these results establish an interface-conscious approach for monolithic integration of REIs with III-V semiconductors, laying the materials foundation for next-generation quantum photonic technologies.

%%%%%%%%%%%%%%%%% Growth Process %%%%%%%%%%%%%%%%%%%%%
\section{Results \& Discussion}

%%%%%%%%%%%%%%%%% Growth Process %%%%%%%%%%%%%%%%%%%%%
\subsection{Growth of Smooth TiO$_2$ Thin Films}
\begin{figure}[!ht]
    \centering
    \includegraphics[width=\textwidth]{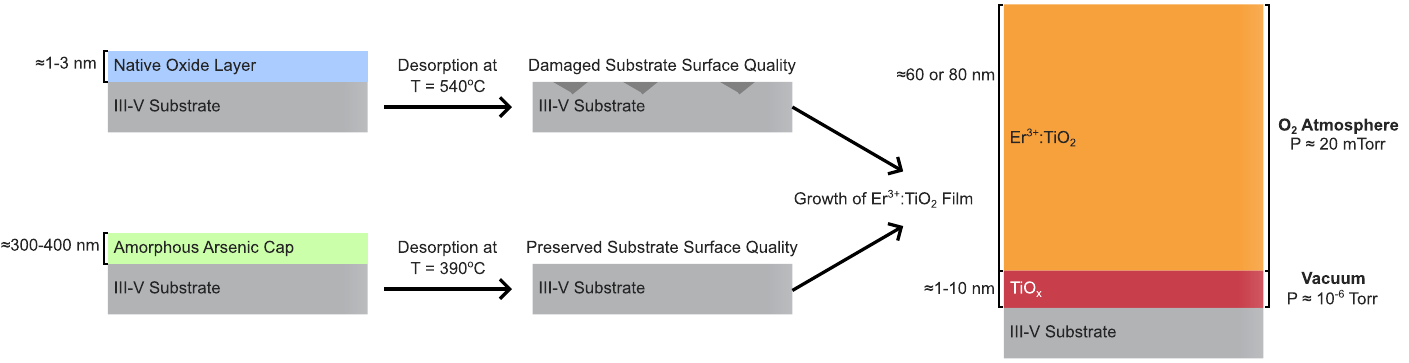}
    \caption{Growth process for bulk-doped  Er$^{3+}$:TiO$_2$-(III-V) samples synthesized using PLD.
    }
    \label{figure:PLD_Growth_Process}
\end{figure}

Er$^{3+}$:TiO$_2$ thin films were synthesized using PLD employing a KrF excimer laser.
A rectangular aperture in a projection beamline defined a quasi-tophat beam profile, enabling a uniform fluence of $2.0~\si{\joule\cdot\centi\meter}^{-2}$ over the illuminated area on the target.
The resulting growth ratewas approximately $0.17 \si{\angstrom}$ per laser shot, as determined from post-growth profilometry and validated by transmission electron microscopy (TEM).
Substrates ($5\times5~\si{\milli\meter}^2$ chips) were mounted on a heated sample holder and the chamber evacuated to high vacuum ($10^{-6}~\si{\torr}$).
As schematized in Figure~\ref{figure:PLD_Growth_Process}, surface preparation of GaAs(100) chips was performed by either thermal desorption of the native oxide ($>540~\si{\celsius}$ \cite{SpringThorpe1987, Tarsa1993, Pun2004, Nosho2002}) or removal of an amorphous arsenic cap ($350-365~\si{\celsius}$ \cite{Resch1992}) deposited using molecular beam epitaxy on a $200~\si{\nano\meter}$ epilayer of GaAs.
Other substrates (GaSb(100), silicon(100)-on-insulator, R-TiO$_2$(110)) underwent only thermal treatment.
Following preparation, a thin, undoped TiO$_2$ buffer layer was deposited under vacuum, and then the oxygen pressure was raised to $20~\si{\milli\torr}$ for the remainder of the growth. 
Each film, $60$–$80~\si{\nano\meter}$ thick, was completed within $\sim20$ minutes, followed by a short (30-minute) oxygen anneal during cooldown.
Within the text, samples are classified according to their substrate and general high or low growth temperature (HT or LT, respectively).

%PLD growth diagram

\begin{figure}[!ht]
    \centering
    \includegraphics[width=\textwidth]{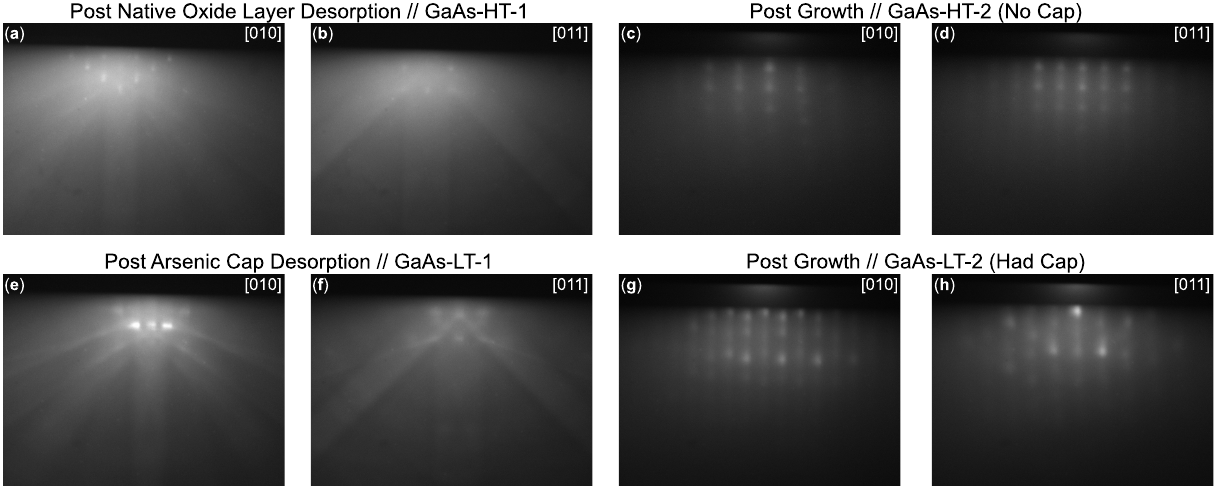}
    \caption{
    Images of the epi-ready GaAs substrate (sample GaAs-HT-1) after native oxide layer desorption are shown in (a) and (b) for the [010] and [011] directions, respectively.
    RHEED patterns for a separate sample (GaAs-HT-2) after $\sim90\si{\nano\meter}$ of TiO$_2$ growth utilizing a similar epi-ready GaAs substrate is shown in (c) and (d) for the [010] and [011] directions, respectively. 
    Images of the GaAs substrate (sample GaAs-LT-1) post-amorphous arsenic cap desorption are shown in (e) and (f) for the [010] and [011] directions, respectively.
    RHEED patterns for a separate sample (GaAs-LT-2) after $\sim60\si{\nano\meter}$ of TiO$_2$ growth also utilizing an amorphous arsenic capped GaAs substrate is shown in (g) and (h) for the [010] and [011] directions, respectively.}
    \label{figure:RHEED_Pre_Growth}
\end{figure}

The crystalline quality of the films was monitored in situ by reflection high-energy electron diffraction (RHEED) at two stages: after substrate surface preparation and following TiO$_2$ film growth.
For oxide-desorbed GaAs (Figure \ref{figure:RHEED_Pre_Growth}(a,b)), the diffraction patterns exhibit weak Kikuchi lines and pronounced spotty features, indicative of a roughened surface and island-like reconstruction reported earlier \cite{Tarsa1993, SpringThorpe1987}.
In contrast, GaAs substrates prepared with an arsenic cap (Figure \ref{figure:RHEED_Pre_Growth}(e,f)) yielded streaky RHEED patterns with sharper Kikuchi lines, confirming the recovery of a smoother and ordered GaAs(100) surface compatible with epitaxial film growth even at low desorption temperatures ($350~\si{\celsius}$).
Films grown on uncapped substrates displayed weaker features consistent with degraded interface quality (Figure \ref{figure:RHEED_Pre_Growth}(c,d)) in comparison to films grown on arsenic-capped GaAs (Figure \ref{figure:RHEED_Pre_Growth}(g,h)), whose post-growth RHEED images of TiO$_2$ exhibited vertical streaks characteristic of predominantly two-dimensional growth and good crystallinity.
The presence of superimposed spots suggests contributions from step edges or islands resulting from buried crystal defects.
%Nevertheless, the relatively weaker intensity of these spots in comparison to streaks between oxide-desorbed samples supports better crystallinity in arsenic-capped samples.
Notably, the period of the RHEED features are significantly different between the two growth processes indicating different crystal structures or phases of the TiO$_2$ thin films.

%TiO2 RHEED Pre Growth

\begin{figure}[h!]
    \centering
    \includegraphics[width=\textwidth]{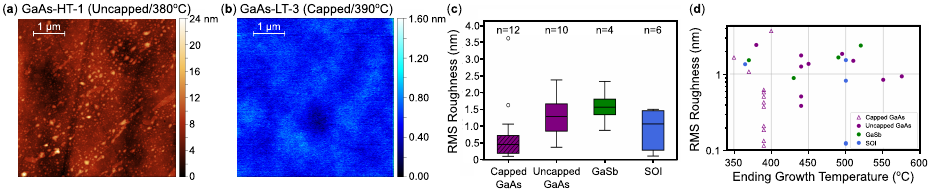}
    \caption{Example AFM scan after post-processing for a (a) low-temperature (A-TiO$_2$)-GaAs sample (GaAs-LT-3) utilizing a protective amorphous arsenic cap and (b) high-temperature (R-TiO$_2$)-GaAs sample (GaAs-HT-1) synthesized after desorbing the native oxide layer present on the GaAs substrate.
    (c) Box-and-whisker plot of mean RMS roughness values extracted across multiple scans organized by substrate.
    (d) Scatter plot comparing ending growth temperature (T$_{grow}$) to mean RMS roughness value extracted across multiple scans.
    }
    \label{figure:AFM}
\end{figure}

Surface morphology was quantified by atomic force microscopy (AFM).
TiO$_2$ films grown after oxide desorption of uncapped GaAs substrates displayed irregular surfaces with some particle agglomeration, yielding root-mean-square (RMS) roughness values exceeding $2~\si{\nano\meter}$ (Figure \ref{figure:AFM}(a)).
In stark contrast, TiO$_2$ films grown on capped GaAs at $T_\textrm{grow}\approx390~\si{\celsius}$ consistently exhibited smooth surfaces with sub-nanometer roughness.
The lowest surface RMS roughness measured over a $5 \times 5~\si{\micro\meter}^2$ area was $116~\si{\pico\meter}$ from the AFM scan shown in Figure~\ref{figure:AFM}(b) (sample: GaAs-LT-3).
The average RMS roughness estimated across several scans was $\approx378~\si{\pico\meter}$ on this sample.

Statistical analysis of surface roughness measured across all samples over a $5 \times 5~\si{\micro\meter}^2$ area is summarized in Figure~\ref{figure:AFM}(c).
Our study confirms that capped GaAs reproducibly yields the smoothest films, while uncapped GaSb and GaAs substrates typically produce rougher surfaces.
Among the TiO$_2$ thin films grown on silicon-on-insulator (SOI), we observed that the addition of a CeO$_2$ buffer at a growth temperature of $500~\si{\celsius}$ improved smoothness, although not with the same consistency as with arsenic-capped GaAs.
In total, 32 samples were analyzed across arsenic-capped GaAs (12), uncapped GaAs (10), uncapped GaSb (4), and silicon-on-insulator (6) substrates, enabling a systematic evaluation of substrate and growth conditions.
We emphasize that not all synthesized films possess a fully Er$^{3+}$-doped layer after the initial undoped buffer layer as depicted in Figure \ref{figure:PLD_Growth_Process}.
In these other cases, the remaining film is either undoped or ``sandwich" doped, that is, only a small $2-10~\si{\nano\meter}$ section of the remaining film, positioned between undoped TiO$_2$ layers, is doped.

At the optimal growth temperature of $390~\si{\celsius}$ on arsenic-capped GaAs, the majority of TiO$_2$ films yielded sub-nanometer RMS roughness ($200-600~\si{\pico\meter}$). 
This reproducibility highlights the robustness of the capping strategy in producing smooth surfaces suitable for nanophotonic integration. 
We note, however, two outliers with RMS roughness above $1~\si{\nano\meter}$ on capped GaAs growth: GaAs-LT-4 ($T_{\text{grow}}=400~^\circ$C) and GaAs-LT-5 ($T_\text{grow}=350~^\circ$C). 
The thin film roughness on GaAs-LT-4 (grown at $400~\si{\celsius}$) exhibited surface contamination likely introduced during transfer to the PLD chamber and the unusually rough surface of GaAs-LT-5 is due to the incomplete desorption of the arsenic cap at the reduced growth temperature ($350~\si{\celsius}$).
Also note that we did not observe a strict correlation between the film RMS roughness and growth temperature as evident in Figure~\ref{figure:AFM}(d) for any substrate, in contrast to previous studies on TiO$_2$ thins film grown on Si(100) substrates using atomic layer deposition \cite{Ji2024}.

%AFM Scans Immediately After Growth

%%%%%%%%%%%%%%%%%  OPTICAL ACTIVATION %%%%%%%%%%%%%%%%%%%%%%%%%%%%%%
\subsection{Optical Activation of Er$^{3+}$ in TiO$_2$ Thin Fims}
\label{section:opt}
A key question for leveraging the Er$^{3+}$:TiO$_2$ thin films in photonic devices is whether the rare-earth ions remain optically active when directly integrated with III-V substrates.
To address this, we combined Raman spectroscopy, which fingerprints the TiO$_2$ crystal phase, with photoluminescence excitation (PLE) spectroscopy of the $Z_1 \rightarrow Y_1$ transition, a sensitive probe of Er$^{3+}$ optical coherence.
Raman spectra were collected at room temperature using a $514~\si{\nano\meter}$ excitation, while PLE was performed at $5\si{\kelvin}$ using a tunable telecom-band laser and time-gated single-photon detection in a confocal microscopy setup.

\label{section:PLE_HighTemp}
%Rutile Raman & Er PLE Measurements
\begin{figure}[h!]
    \centering
    \includegraphics[width=\textwidth]{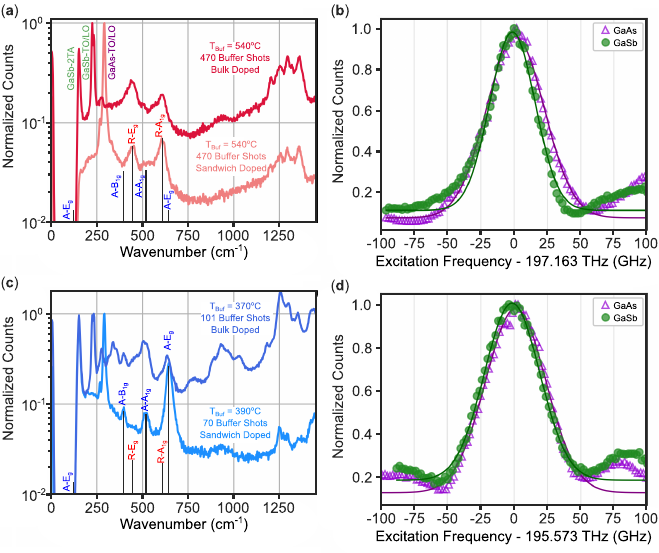}
    \caption{(a) Raman spectroscopy results for two high temperature (R-TiO2)-III-V samples.
    The dark red trace is sample GaSb-HT-1 (bulk-doped; GaSb substrate), and the light red trace is sample GaAs-HT-3 (sandwich-doped; GaAs substrate).
    Sample buffer shots and buffer growth temperatures are included within the figure.
    (b) $Z_1 \rightarrow Y_1$ Er$^{3+}$ PLE results for samples GaAs-HT-4 (purple) and GaSb-HT-1 (green).
    (c) Raman spectroscopy results for two low temperature (A-TiO2)-III-V samples.
    The dark blue trace is sample GaSb-LT-1 (bulk-doped; GaSb substrate), and the light blue trace is sample ST2417 (sandwich-doped; GaAs substrate).
    Sample buffer shots and buffer growth temperatures are included within the figure.
    (d) $Z_1 \rightarrow Y_1$ Er$^{3+}$ PLE results for samples GaAs-LT-1 (purple) and GaSb-LT-1 (green).
    }
    \label{figure:RamanPLE}
\end{figure}

Across the Er$^{3+}$:TiO$_2$ thin film samples, two distinct optical signatures emerged, correlating with substrate preparation and growth conditions.
Films synthesized at elevated substrate temperatures ($\geq450~\si{\celsius}$) on oxide-desorbed GaAs or GaSb crystallized in the rutile phase (R-TiO$_2$), with Raman spectra showing the characteristic E$_g$ ($449~\si{\centi\meter}^{-1}$) and A$_{1g}$ ($614~\si{\centi\meter}^{-1}$) modes (Figure~\ref{figure:RamanPLE}(a)). 
In erbium-doped samples, additional features around $1300~\si{\centi\meter}^{-1}$ correspond to visible fluorescence from Er$^{3+}$ ions under $514~\si{\nano\meter}$ excitation.
Specifically, the $4f$-electrons are excited into the $^{4}$S$_{3/2}$ or $^{2}$H$_{11/2}$ manifolds, which relax via a combination of phonon-assisted and radiative processes to the $^{4}$I$_{15/2}$ ground state, producing visible fluorescence around $550~\si{\nano\meter}$ ($\sim1300~\si{\centi\meter}^{-1}$).
This confirms Er$^{3+}$ incorporation; however, the fluorescence is strongly phonon-coupled and unsuitable for linewidth or coherence analysis. 
Cryogenic PLE measurements revealed a sharp crystal-field split $Z_1 \rightarrow Y_1$ resonance of the $^{4}$I$_{13/2} \rightarrow$ $^{4}$I$_{15/2}$ transition at $197.16 ~\si{\tera\hertz}$ $(1520.5~\si{\nano\meter})$ (Figure~\ref{figure:RamanPLE}(b)), consistent with \emph{ab initio} predictions of Er$^{3+}$ substituting at Ti$^{4+}$ sites in R-TiO$_2$ lattice \cite{limbu2025}.
On GaAs substrates, we measured an inhomogeneous linewidth of $50.9(7) ~\si{\giga\hertz}$ and a lifetime of $5.3(3) ~\si{\milli\second}$, while on GaSb the linewidth narrowed to $40(1) ~\si{\giga\hertz}$ but the lifetime shortened to $4.7(2) ~\si{\milli\second}$. 
These inhomogeneous linewidths are comparable to prior results for highly-doped ($>1000$ppm) R-TiO$_2$ thin films grown on silicon \cite{Martins2024,Singh2024}.
The measured lifetimes, while presenting substrate-dependent variations, are similar to prior theoretical \cite{Dodson2012} and experimental results \cite{Phenicie2019}, demonstrating that direct III-V integration preserves Er$^{3+}$ optical activity.

At lower growth temperatures ($370-390~\si{\celsius}$) on arsenic-capped GaAs or oxide-desorbed GaSb, the films stabilized in the anatase phase (A-TiO$_2$).
Raman spectra revealed the characteristic B$_{1g}$ ($\sim 399~\si{\centi\meter}^{-1}$), A$_{1g}$ ($\sim 515~\si{\centi\meter}^{-1}$), and E$_g$ ($\sim 639~\si{\centi\meter}^{-1}$) modes, along with a pronounced low-frequency E$_g$ peak ($\sim 144~\si{\centi\meter}^{-1}$), indicative of good crystallinity (Figure~\ref{figure:RamanPLE}(c)).
Cryogenic PLE spectra of these films showed the $Z_1 \rightarrow Y_1$ resonance at $195.57 ~\si{\tera\hertz} ~(1532.9~\si{\nano\meter})$, distinctly shifted from the R-TiO$_2$ films indicating Er$^{3+}$ substituting at Ti$^{4+}$ sites in A-TiO$_2$ lattice (Figure~\ref{figure:RamanPLE}(d)). 
On GaAs, the inhomogeneous linewidth was $53(1) ~\si{\giga\hertz}$ with a lifetime of $1.70(8) ~\si{\milli\second}$, while on GaSb the linewidth narrowed to $48.9(6) ~\si{\giga\hertz}$ but the lifetime shortened to $1.34(7) ~\si{\milli\second}$. 
Compared to rutile, the anatase films exhibited significantly shorter lifetimes, consistent with earlier reports \cite{Ji2024, Singh2024}.

\begin{figure}[h!]
    \centering
    \includegraphics[width=0.5\textwidth]{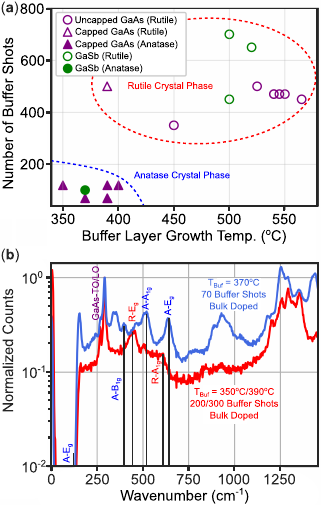}
    \caption{(a) Phase diagram for all synthesized TiO$_2$ thin films on GaAs and GaSb substrates.
    (b) Raman spectra for the open purple triangle R-TiO$_2$ film on GaAs substrate (red trace; sample GaAs-LT-6) data point in (a) along with a bulk-doped A-TiO$_2$ film on GaAs substrate (blue trace; sample GaAs-LT-1) for comparison.
    Relevant phonon modes with drop lines to guide the eye are labeled for GaAs, A-TiO$_2$ (blue ``A" prefix), and R-TiO$_2$ (red ``R" prefix).}
    \label{figure:Phase_Diagram}
\end{figure}

The phase selectivity of TiO$_2$ is not determined by growth temperature alone but also by buffer-layer thickness, as summarized in the phase diagram (Figure~\ref{figure:Phase_Diagram}(a)).
Notably, two capped-GaAs films grown at $390~\si{\celsius}$, but with different buffer thicknesses, yielded distinct phases: a thinner buffer (70 laser shots) stabilized anatase phase, whereas a thicker buffer (500 laser shots) drove rutile formation.
The Raman spectra (Figure~\ref{figure:Phase_Diagram}(b) clearly illustrate this contrast with the thick-buffer sample showing a complete suppression of the $144~\si{\centi\meter}^{-1}$ E$_g$ mode.
This anomalous growth highlights how strain relaxation and oxygen-vacancy accumulation beyond a critical buffer thickness can tip the balance toward rutile even at reduced growth temperatures. 
Such sensitivity underscores the importance of buffer engineering in controlling phase stability and, by extension, Er$^{3+}$ optical activation.

%Phase Diagram

Interestingly, in both rutile and anatase films, samples grown on GaSb reproducibly exhibited narrower, inhomogeneous linewidths, yet shorter optical lifetimes, than those on GaAs. 
Such decoupling of linewidth and lifetime has been observed in other REI-doped oxides, where ensemble inhomogeneity is dominated by static disorder (strain fields, compositional fluctuations), while the homogeneous linewidth and lifetime are set by dynamic decoherence and nonradiative processes (spin flips, spectral diffusion, or defect-assisted relaxation) \cite{Ferrier2013, Fossati2023}.
Similarly, photon-echo studies demonstrated that magnetic-noise-induced spectral diffusion can broaden homogeneous linewidths without altering ensemble disorder \cite{Bottger2009, Thiel2011}.
In our III-V integrated films, the systematically shorter lifetimes on GaSb are therefore likely linked to interface-specific nonradiative pathways, including Ga diffusion into the oxide lattice \cite{VanOmmen1987, Ye2015}, differences in Sb- versus As-terminated interface chemistries \cite{Zinck1997}, and oxygen-vacancy-mediated quenching \cite{Tuller2011, Linderlv2020}, which can enhance decay rates without substantially increasing the ensemble disorder.
Overall, Er$^{3+}$ remains optically active in both phases on III-V substrates with A-TiO$_2$, grown at a lower temperature with a smoother morphology, providing a suitable platform for low-loss nanophotonics. 
This motivates the structural analysis below to pinpoint which microstructural and interfacial defects set the observed lifetime/linewidth tradeoffs. 

In contrast to previous growths of TiO$_2$ on GaAs \cite{Liu2001_1, Liu2001_2}, we can additionally synthesize A-TiO$_2$ on GaAs by utilizing two separate oxygen pressures during growth.
Furthermore, we have demonstrated the first-reported synthesis of both R-TiO$_2$ and A-TiO$_2$ on GaSb, an emerging platform for telecom-band single-photon sources.

%%%%%%%%%%%%%%%%%%% STRUCTURAL ORIGINS %%%%%%%%%%%%%%%%%%%%%%%%

%%%%%%%%%%%%%%%%%%%%%%%%% Microstructure %%%%%%%%%%%%%%%%%%
\subsection{Crystallographic Phase and Microstructure Analysis}
The crystallographic orientation and microstructure of the TiO$_2$ films were investigated using minimum coincident interface area (MCIA) modeling in conjunction with $\theta-2\theta$ X-ray diffraction (XRD) measurements.
MCIA provides a geometric metric for predicting orientation-selective epitaxy by quantifying the smallest lattice-commensurate overlap between a film and substrate \cite{Zur1984, Ding2016}, while XRD directly proves the resulting out-of-plane order, grain size, and strain.
Together, these methods establish the connection between interface energetics and the observed structural phase.
Note that because MCIA assumes atomically sharp, defect-free interfaces, which are challenging to realize in PLD, it serves as a predictive metric rather than a guarantee.
In practice, maintaining a high-quality interface, such as through arsenic capping, is essential for approaching the geometric minimum.

%MCIA maps
\begin{figure}[h!]
    \centering
    \includegraphics[width=0.5\textwidth]{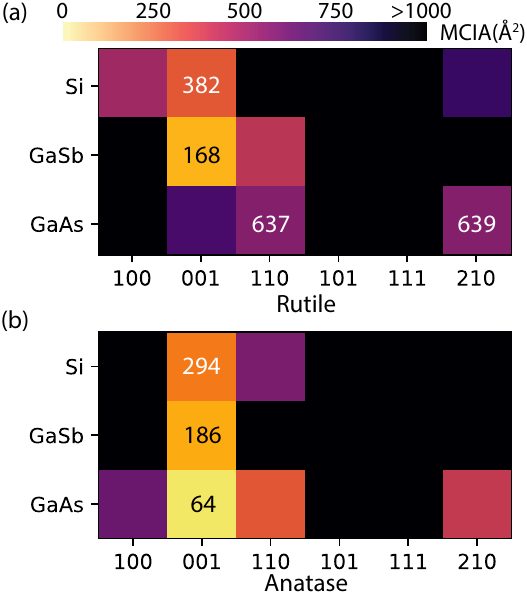}
    \caption{Calculated minimal co-incident interface area (MCIA) for (100)-oriented Si, GaSb, and GaAs substrates with different crystal orientations, labeled by the corresponding Miller index, of (a) rutile and (b) anatase phase TiO$_2$. The smallest MCIA values for a given substrate-film pair are marked in the plot.}
    \label{fig:MCIAmaps}
\end{figure}

The MCIA maps in Figure \ref{fig:MCIAmaps} reveal clear orientation-dependent trends consistent with the experimentally observed phase selectivity.
For GaAs, anatase (001) exhibits the smallest MCIA value ($64~\si{\angstrom}^2$), nearly an order of magnitude lower than rutile (110).
Such a small MCIA value suggests that even moderate interfacial disorder or step-edge roughness can still support anatase-phase epitaxy at lower growth temperatures, consistent with surface-energy minimization predicted by MCIA.
At higher temperatures ($>450~\si{\celsius}$), enhanced adatom mobility and oxygen incorporation promote atomic rearrangement toward bulk-energy minimization, favoring the thermodynamically denser rutile phase.
However, rutile (110) and (210) are nearly degenerate in MCIA, accounting for the mixed-texture films observed at elevated growth temperatures \cite{Liu2001_1, Liu2001_2}.
For GaSb, both anatase (001) and rutile (001) orientations yield comparably small MCIA values ($\approx 175~\si{\angstrom}^2$), implying that the energy balance between strain and chemical bonding, rather than pure lattice matching, determines which phase forms.
By comparison, the anatase and rutile phase MCIA values on Si (100) are comparable ($300~\si{\angstrom}^2$), consistent with the polycrystalline TiO$_2$ films typically reported on silicon \cite{Singh2024,Ji2024}.
These geometric trends provide a predictive framework linking interface geometry to phase selectivity observed in Figure~\ref{figure:Phase_Diagram}, motivating the experimental validation of crystal orientation using XRD.

%R-TiO2 and A-TiO2 on III-V XRD Spectra
\begin{figure}[h!]
    \centering
    \includegraphics[width=\textwidth]{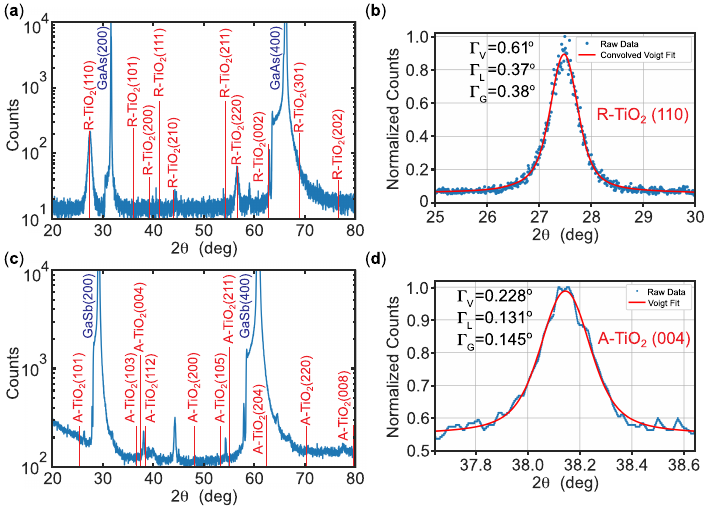}
    \caption{(a)-(b) XRD spectra for sample GaAs-HT-2.
    (a) Wide $\theta$-$2\theta$ scan of the sample.
    Vertical red lines indicate the expected R-TiO$_2$ peak positions of a perfect crystal using a CuK$_{\alpha 1}$ x-ray source.
    (b) Zoom-in of the R-TiO$_2$ (110) XRD peak and its resulting convolved Voigt fit.
    (c)-(d) XRD results for the (A-TiO$_2$)-GaSb sample (GaSb-LT-1).
    (c) $\theta$-$2\theta$ XRD spectra.
    Vertical red lines indicate the expected R-TiO$_2$ peak positions of a perfect crystal using a CuK$_{\alpha 1}$ x-ray source.
    (d) Zoom-in of the A-TiO$_2$ (004) XRD peak and its resulting Voigt fit.
    }
    \label{figure:XRD_Panel}
\end{figure}

A representative $\theta$-2$\theta$ XRD scan of a TiO$_{2}$ thin film synthesized on oxide-desorbed GaAs at high growth temperatures ($565~\si{\degree}$C) is shown in Figure~\ref{figure:XRD_Panel}(a).
The dominant reflection peak at $2\theta = 27.4\si{\degree}$ matches the rutile (110) plane, in agreement with MCIA predictions.
The corresponding (220) harmonic and a weak (210) reflection are also observed, with the relative intensities suggesting a partial preferred orientation rather than a fully random polycrystalline structure.
Using the Scherrer and Wilson equations upon fitting the (110) peak to a Voigt lineshape (Figure~\ref{figure:XRD_Panel}(b)), the extracted grain size ($\tau$) and microstrain ($\epsilon$) were determined to be $22\pm1~\si{\nano\meter}$ and $0.68 \pm 0.04\%$, respectively.
Given the estimated film thickness of $\sim$90~nm for high-temperature growths, the $\sim22~\si{\nano\meter}$ grain size confirms that the R-TiO$_2$ films are polycrystalline, consistent with prior reports \cite{Liu2001_1, Liu2001_2}.
The nonzero $\epsilon$ indicates that even after grain breakup, a residual tensile component remains within the film.
The overall trend of small, tensile-strained crystallites is consistent across all high-temperature samples, with $\tau$ and $\epsilon$ ranging from $14-32~\si{\nano\meter}$ and $0.47-0.96\%$, respectively.
These microstructural characteristics--polycrystallinity, partial preferred orientation, and residual tensile strain--are consistent with the high and near-degenerate MCIA values predicted for (110)/(210) R-TiO$_2$ orientations on GaAs.

At lower growth temperatures ($<390~\si{\celsius}$), the XRD signal from TiO$_2$ grown on arsenic-capped GaAs was too weak to unambiguously resolve diffraction peaks under standard $\theta$–2$\theta$ geometry, likely due to the film’s limited thickness ($\approx 60~\si{\nano\meter}$) and low scattering volume \cite{Guillen2025}.
Nevertheless, the smooth AFM morphology and distinct Raman signatures (Figure~\ref{figure:Phase_Diagram}) indicate crystalline anatase formation.
Grazing-incidence XRD measurements further revealed a weak but reproducible A-TiO$_2$ (101) reflection, providing additional confirmation of the anatase phase despite the limited film thickness.
A measurable XRD signal was also obtained for an analogous film grown on GaSb in the same temperature range (Figure~\ref{figure:XRD_Panel}(c)).
A narrow A-TiO$_2$(004) reflection is evident (Figure~\ref{figure:XRD_Panel}(d)), confirming anatase-phase stabilization under diffusion-limited growth conditions.
Voigt-profile fitting yields a grain size of $64(2)~\si{\nano\meter}$, comparable to the film thickness, and a microstrain of $\epsilon=0.183(6)\%$, suggesting near-epitaxial growth with minimal mosaicity.
The fitted Voigt linewidth is nearly a third of the high-temperature rutile film on GaAs (Figure~\ref{figure:XRD_Panel}(b)), consistent with a substantial reduction in microstrain broadening and overall improvement in crystalline quality.
The extracted A(004) peak center of $38.144(1)\si{\degree}$, corresponds to a lattice parameter $c=9.43(1)~\si{\angstrom}$, indicating a minor compressive strain consistent with the interfacial registry predicted by MCIA.
Overall, the transition from polycrystalline, tensile-strained rutile to smooth, low-strain anatase with decreasing growth temperature aligns with the MCIA predictions and thermodynamic trends discussed earlier.

%%%%%%%%%%%%%%%%%%%%%%%% Interface Chemistry & Defects %%%%%%%%%%%%%%%%%%%
\subsection{Interface Chemistry \& Defects}
Building on the structural and phase evolution described above, we next examine the atomic-scale interface chemistry and defect structure that govern TiO$_2$-(III-V) heteroepitaxy.
The transition from tensile-strained rutile to relaxed anatase, together with the sensitivity to substrate termination, indicates that interfacial bonding and stoichiometry critically determine both phase stability and optical performance.
To elucidate these effects, we performed cross-sectional transmission electron microscopy (TEM) and electron energy-loss spectroscopy (EELS) on a sandwich-doped oxide-desorbed GaAs sample with an R-TiO$_2$ film (sample GaAs-HT-5).
This sample contained a $10~\si{\nano\meter}$ oxygen-deficient TiO$_\textrm{x}$ buffer grown at $545~\si{\celsius}$, followed by an 80~\si{\nano\meter} TiO$_2$ layer deposited at $20~\si{\milli\torr}$ oxygen pressure.
A central $10~\si{\nano\meter}$ region of the film was selectively doped with erbium using a 3000 ppm Er$^{3+}$:TiO$_2$ target, intended to trace possible Er$^{3+}$ migration along the growth detection.
Although this sample represents a single high-temperature growth, it serves as a detailed case study for understanding interface reactions and defect formation in the rutile regime.

%TiO2 Large FOV TEM Image Panel
\begin{figure}[h!]
    \centering
    \includegraphics[width=\textwidth]{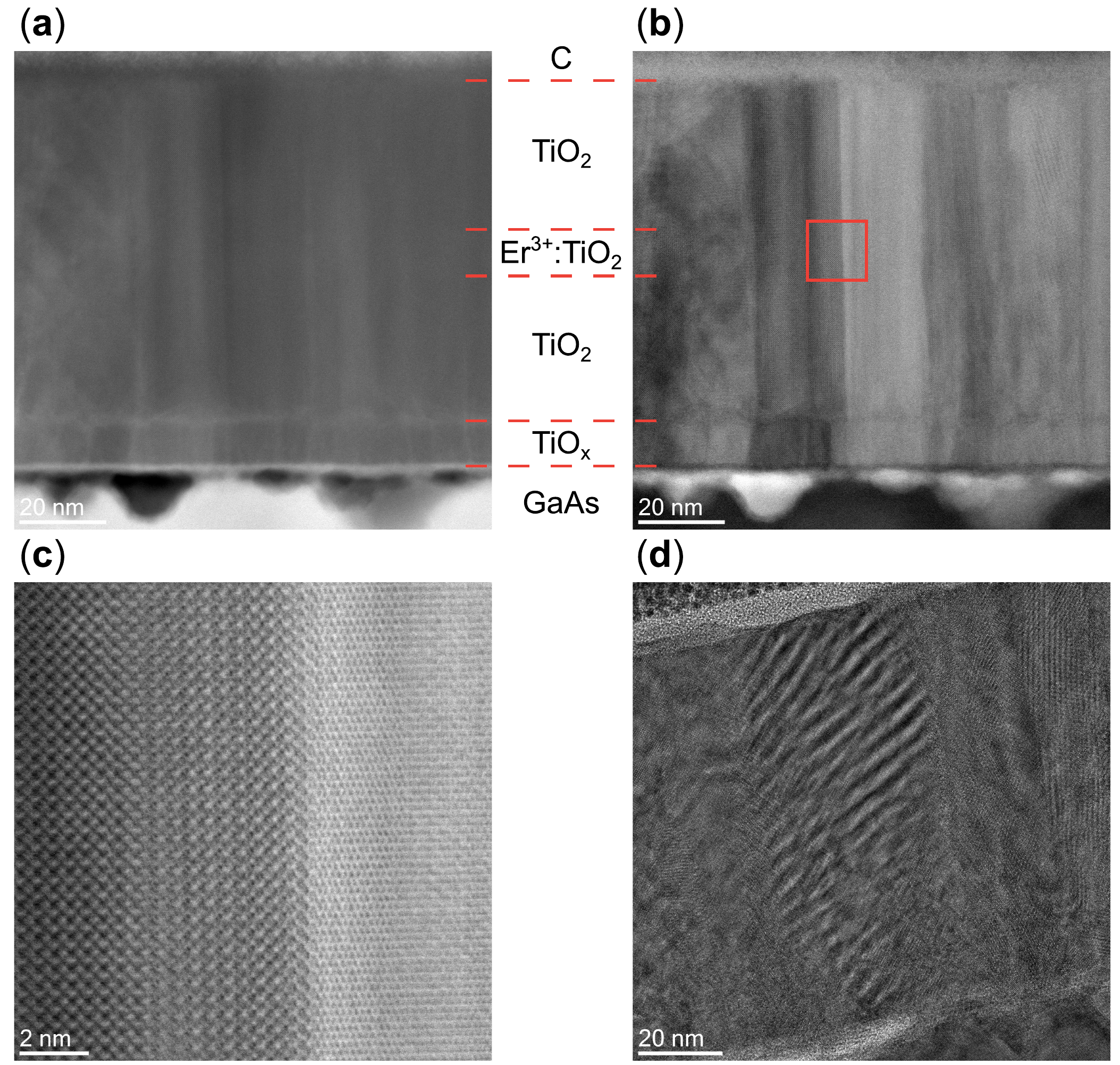}
    \caption{TEM images of the (Er$^{3+}$:R-TiO$_2$)-GaAs sample. (a) HAADF-STEM image showing the columnar structure of the TiO$_{2}$ film. 
    (b) Complementary LAADF-STEM image further highlighting the film's composition as a function of growth position with dashed lines.
    (c) The inset from (b) shows the polycrystalline nature of the film and the atomic-level transition between grains. 
    (d) HRTEM image of the TiO$_{2}$ thin film. Moir\'{e} fringes where the crystal domains twist and overlap are clearly visible and extend over several tens of nanometers in the growth direction.}
    \label{figure:TEM}
\end{figure}

Scanning transmission electron microscopy (STEM) imaging was utilized to study the film morphology. Both high-angle and low-angle annular dark field (HAADF and LAADF, respectively) STEM were employed (Figure~\ref{figure:TEM}(a,b)). 
HAADF contrast is well-established to originate primarily from atomic mass, with heavier atoms appearing brighter, whereas LAADF is largely influenced by diffraction contrast \cite{Phillips2012}. 
This manifests as the HAADF image (Figure~\ref{figure:TEM}(a)) providing directly interpretable contrast information (relating primarily to atomic number with minimal diffraction contributions) whereas LAADF imaging (Figure~\ref{figure:TEM}(b)) highlights the presence of inter-column crystal rotation.
The GaAs substrate exhibits dark/bright pits and an uneven surface topography consistent with the oxide-desorption-induced damage.
High-resolution TEM (HRTEM) and STEM images of these nanoscale depressions indicate gallium-deficient voids, where disrupted surface reconstruction locally alters the nucleation density and promotes nonuniform columnar alignment of the TiO$_2$ film.
The oxygen-deficient buffer can be distinguished from the overlying TiO$_2$ layer by a discontinuity in the columnar texture.
The high-resolution inset from Figure~\ref{figure:TEM}(b) (red box) shown in Figure~\ref{figure:TEM}(c) reveals twisting and reorientation of atomic planes between neighboring regions, indicating the presence of misoriented crystal domains throughout the film. 
The Moir\'{e} fringes observed throughout the wide-view HRTEM image in Figure~\ref{figure:TEM}(d) arise from this crystal misorientation when viewed in projection. 
These fringes extend through the entire film thickness, reflecting the complex rotational variants occurring on the few-nanometer scale.
Similar patterns were observed at multiple regions across the cross-section, confirming that crystal misorientation is a pervasive microstructural feature. The frequent observation of both in-plane and out-of-plane rotations within the <100 nm thick lamella suggests the presence of small grains throughout the film.

%TiO2 ABF Image
\begin{figure}[h!]
    \centering
    \includegraphics[width=\textwidth]{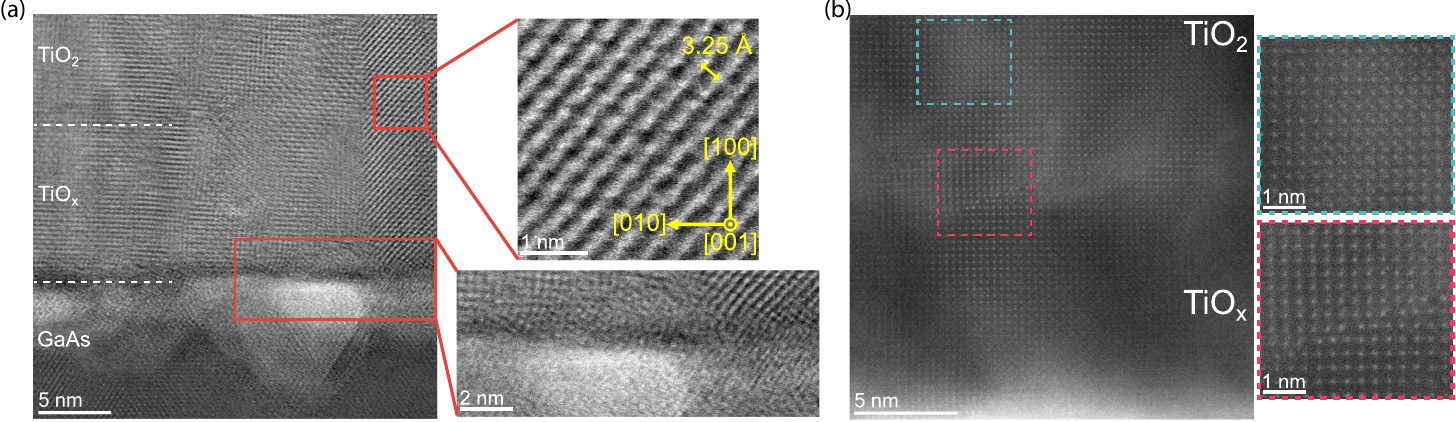}
    \caption{(a) Annular bright-field (ABF) image of the TiO$_x$/TiO$_2$/GaAs interface. 
    Crystals grown at the undamaged GaAs substrate sites promote (110) R-TiO$_{2}$ growth as shown in the top inset. 
    The bottom inset shows a zoom-in of the damaged interface, and more specifically, a characteristic (gallium-deficient) pit (bright region) within the GaAs substrate and the beginning few monolayers of the TiO$_{x}$ buffer layer. 
    (b) HAADF-STEM image showing the atomic-level resolution of the TiO$_{x}$/TiO$_{2}$ interface.
    The blue inset shows an additional plane of atoms resulting from the rotation/twisting of the local crystal domain.
    The red inset shows the immediate surroundings of multiple vacancy centers, which cause lattice dislocations.}
    \label{figure:TEM_Interface}
\end{figure}

Atomic-resolution STEM imaging of the interface (Figure~\ref{figure:TEM_Interface}) further clarifies the origin of these columnar and Moir\'{e} textures.
A $\approx5~\si{\nano\meter}$ region exhibiting varying contrast is evident within the GaAs substrate, as shown in Figure~\ref{figure:TEM_Interface}(a). 
The surface topography and local substrate orientation directly affect the quality of the initial TiO$_\textrm{x}$ monolayers: regions above the pits appear irregular, while adjacent areas nucleate ordered rutile (110) crystallites that extend through the film thickness.
The measured $3.25~\si{\angstrom}$ lattice spacing of these (110) planes matches bulk R-TiO$_2$ values, suggesting that strain release occurs rapidly during early growth.
A narrow ($1-2~\si{\nano\meter}$) transition zone separates the TiO$_\textrm{x}$ buffer from the subsequent TiO$_2$ layer, marking the point where oxygen pressure was increased during deposition, which we probe further using atomic-resolution HAADF-STEM imaging in Figure~\ref{figure:TEM_Interface}(b).

Two characteristic features are apparent in this transition region, marked by the dashed boxes.
The blue inset reveals multiple crystal planes that are slightly misoriented from the main lattice, consistent with the local rotation of crystal domains that produces Moir\'{e} fringes.
The red inset highlights multiple lattice dislocations and accompanying vacancies, with a few atomic columns displaying enhanced contrast and irregular spacing, potentially corresponding to isolated gallium atoms incorporated during interdiffusion.
Although Er$^{3+}$ ions could not be directly resolved within the 10 nm doped region, this is expected given their low concentration.
Future STEM measurements on films with higher doping concentrations and enhanced signal-to-noise ratio may enable quantitative analysis of Er-O bond lengths and local substitution environments, including the influence of nearby oxygen vacancies.

%%%%%%%%%%%%%%%%% Electron Energy Loss Spectroscopy %%%%%%%%%%%%%%%%%%%%%
%TiO2 EELS
\begin{figure}[h!]
    \centering
    \includegraphics[width=0.5\textwidth]{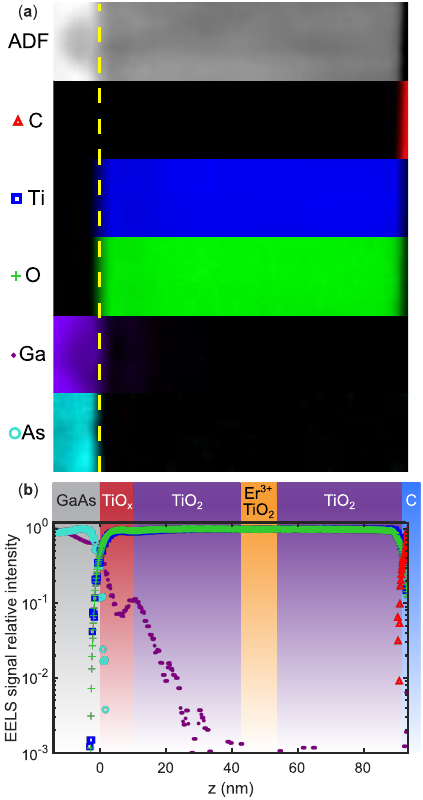}
    \caption{(a) Relative EELS signal intensity of each atom composing the TiO$_{2}$/GaAs interface as a function of position from the GaAs substrate.
    (b) Integrated EELS signal of each element composing the TiO$_{2}$/GaAs interface.
    Each elemental signal displayed is normalized to its column of pixels with the most integrated counts within each respective EELS map in (a).
    The carbon signal atop the TiO$_2$ originates from the protective layer used during ion beam preparation of the TEM lamella.}
    \label{figure:EELS}
\end{figure}

To quantify the compositional gradients observed in the STEM imaging, we performed electron energy-loss spectroscopy (EELS) mapping across the same lamella (Figure~\ref{figure:EELS}(a)).
The elemental distributions of Ga, As, Ti, and O delineate a sharp transition between the GaAs substrate and the TiO$_2$ film, while the Er$^{3+}$ signal remained below the detection threshold due to its low concentration.
Depth-integrated line profiles, normalized to the maximum signal for each element (Figure~\ref{figure:EELS}(b)), reveal that Ti and O intensities increase abruptly at the interface, whereas Ga exhibits pronounced depletion on the substrate side accompanied by correlated diffusion into the oxide film.
A localized accumulation of Ga occurs near the TiO$_\textrm{x}$/TiO$_2$ boundary, corresponding to the region where growth was briefly paused during the oxygen-pressure ramp.
This Ga migration behavior suggests that cation interdiffusion competes with oxygen incorporation during the transition from oxygen-deficient to stoichiometric growth. 
This process produces a thin compositionally-graded zone enriched with Ga defects, which likely seeds the non-uniform columnar nucleation observed in STEM. 
Fitting the depth-resolved Ga profiles yields characteristic diffusion lengths of $4~\si{\nano\meter}$ and diffusion coefficients on the order of $10^{-17}~\si{\centi\meter}^2\si{\second}^{-1}$, consistent with low-activation-energy Ga diffusion in other oxide materials \cite{Katz1969, Wagner1974, VanOmmen1987}.

%%%%%%%%%%%%%%%%% Oxygen Vacancies in TiO$_2$ Films %%%%%%%%%%%%%%%%%%%%%
Although the oxygen EELS signal varies with depth, the quality of the lamella in this region prevents unambiguous separation of oxygen- and titanium-vacancy contributions near the interface.
Beyond this region, a slight reduction in the oxygen signal relative to titanium indicates a finite population of residual oxygen vacancies that persist even after deposition and cooldown in the $20~\si{\milli\torr}$ oxygen pressure.
To qualitatively assess their role, we modeled vacancy incorporation, diffusion, and annihilation during the oxygen-pressure ramp using a simple rate-equation framework \cite{Tuller2011}.
The simulations show that once the oxygen-deficient buffer exceeds a critical thickness, the average oxygen vacancy concentration saturates, limiting reoxidation even after extended oxygen exposure.
This residual vacancy population provides a plausible pathway for strain relaxation and polycrystalline rutile nucleation at lower growth temperatures, consistent with the phase diagram and microstructural observations above.
Conceptually, the model supports a picture in which oxygen-vacancy accumulation couples interfacial chemistry to strain relief and phase stability in TiO$_2$-(III-V) heterostructures. 

% %Buffer Layer V_O Diffusion
% \begin{figure}[h!]
%     \centering
%     \includegraphics[width=0.5\textwidth]{Figures/Vacancy_Diffusion_Buffer_Layer_Thickness_Dependence.pdf}
%     \caption{Average $V_{O}$ density, calculated by taking the mean of the depth-resolved $V_{O}$ density, for various buffer layer shots before and after O$_2$ purge.}
%     \label{figure:Vacancy_Diffusion_Buffer}
% \end{figure}

\subsection{Linking Microstructure to Optical Activation}
The microstructural and interfacial trends described above provide a physical context for the optical activity of Er$^{3+}$ discussed in Sec.~\ref{section:opt}.
Residual strain and oxygen vacancies can each contribute to the inhomogeneous linewidths by locally perturbing the crystal-field splitting experienced by individual erbium ions \cite{Kunkel2016prb, Fossati2023, Brinn2025}.
In addition, diffusion of Ga atoms into the oxide introduces excess nuclear-spin noise that broadens the optical linewidth and reduces the spin-coherence time of Er$^{3+}$ ions \cite{Jamonneau2016prb, Choi2017prl, Candido2024}.
In the R-TiO$_2$ films, where partial strain relaxation occurs, the broader inhomogeneous PLE linewidths observed on GaAs ($50~\si{\giga\hertz}$) relative to GaSb ($40~\si{\giga\hertz}$) likely reflect differences in interfacial strain accommodation arising from the significant difference in the corresponding MCIA values.
By contrast, the nearly identical linewidths observed in the A-TiO$_2$ films suggest that interfacial strain is not the dominant factor.
Instead, static disorder associated with oxygen vacancies and defect complexes, more readily formed in the anatase phase, becomes the primary source of dephasing. 
However, the uniform doping across the film thickness employed in the samples shown in Figure~\ref{figure:RamanPLE} makes it difficult to decouple the effects of substrate-induced local strain and nuclear (Ga atoms) or electronic (O vacancies) spin noise on the rare-earth ions.
A more detailed study employing spatially engineered Er$^{3+}$ doping profiles, which allow selective probing of ions at controlled distances from the interface, is therefore required to separate these effects quantitatively and represents an important direction for future work.

%%%%%% Conclusion %%%%%%
\section{Conclusion}
Understanding and controlling the interplay between microstructure, interface chemistry, and optical activation is central to realizing coherent rare-earth emitters integrated with semiconductor photonic platforms.
In this work, we established how substrate termination, oxygen stoichiometry, and strain collectively govern phase selection and defect formation in TiO$_2$-(III-V) heterostructures.
Cross-sectional STEM-EELS analysis revealed interfacial Ga diffusion and oxygen-vacancy accumulation promote strain relaxation and rutile phase nucleation. 
In contrast, low-defect, arsenic-capped substrates that preserve interfacial integrity favor epitaxial anatase growth at low growth temperatures.
These microscopic processes directly correlate with the optical response of Er$^{3+}$:TiO$_2$, where the balance between strain relaxation and defect-induced disorder determines the inhomogeneous linewidths and optical coherence.
The mechanistic insight gained here thus links growth thermodynamics to optical activation, providing a framework for engineering oxide-semiconductor heterostructures for integrated photonics.

Further improvements in film quality and stoichiometry could be achieved through precise control of plasma plume kinetics and oxygen chemical potential during growth \cite{Lee2016}.
Introducing a diffusion barrier in tandem with arsenic capping the substrate for preserving surface integrity could help minimize interfacial pits and mitigate gallium interdiffusion.
Finally, depth-selective rare-earth doping could enable targeted probing of strain and defect effects on optical coherence, providing a pathway to disentangle microscopic noise sources.
Collectively,  these refinements represent critical steps toward scalable, telecom-compatible, rare-earth-doped oxide films monolithically integrated with III-V semiconductors for on-chip quantum photonic technologies.

%%%%%%%%%%%%%%%%%%%%%%%%%  METHODS %%%%%%%%%%%%%%%%%%%%%%%%%%
\section{Methods}

\subsection*{Target Preparation}
One-inch diameter undoped and 3000 ppm Er$^{3+}$ doped TiO$_2$ targets were fabricated from TiO$_2$ and Er$_2$O$_3$ powders (Sigma Aldrich) by cold-pressing and high-temperature sintering ($1600~\si{\celsius}$).

\subsection*{Surface and Morphology Characterization}
RHEED patterns were measured in situ with an electron beam operated at $20~\si{\kilo\volt}$ with $1.4~\si{\milli\ampere}$ emission current and a phosphor screen. 
Surface morphology was measured by tapping-mode atomic-force microscopy (Bruker Dimension Icon) equipped with a Si probe (TESPA-V2, $7~\si{\nano\meter}$ tip radius, $37~\si{\newton/\meter}$ spring constant) and the data analyzed using Gwyddion \cite{Necas2012}. 
Crystallographic phase and orientation were identified by X-ray diffraction (Rigaku SmartLab or a four-circle Panalytical X'Pert), with peaks fit to a Voigt-function to estimate the average grain size and microstrain in the film.

\subsection*{Transmission Electron Microscopy}
Cross-sectional lamellae were prepared using a Xe$^{+}$ plasma focused ion beam (Helios~5~UXe, Thermo Fisher Scientific), which avoids Ga implantation and amorphization effects common to conventional Ga FIB milling in III-V materials. 
The extracted lamellae were analyzed using a probe- and image-corrected STEM (Spectra Ultra, Thermo Fisher Scientific) equipped with an X-FEG/UltiMono source operating at $300~\si{\kilo\volt}$ accelerating voltage.
Images were acquired using a $28~\si{\milli\radian}$ convergence semi-angle with $\sim 110$--$130~\si{\pico\ampere}$ beam current. 
Acceptance angles for the ABF, LAADF, MAADF, and HAADF detectors were 0--11, 12--23, 23--44, and 49--200 mrad, respectively. 
HRTEM images were acquired using parallel illumination on a Ceta-S detector.
STEM-EELS data were acquired with a ContinuumK3 (Gatan) using the spectrometer's secondary detector (fiber-optically coupled scintillator, model 1069.EXUP).
Spectrum images were acquired in DualEELS mode using a $53~\si{\milli\radian}$ collection semi-angle and $0.3~\si{\electronvolt}$/channel dispersion. 

\subsection*{Optical Characterization}
Room-temperature Raman spectroscopy measurements were performed using a dispersive Raman spectrometer (Renishaw) with a laser wavelength of $\lambda = 514~\si{\nano\meter}$ and a 50$\times$ microscope objective.
A long-pass filter blocked the laser and transmitted signal with a Stokes shift of at least $180~\si{\centi\meter}^{-1}$ Raman shift.
Cryogenic photoluminescence-excitation (PLE) spectroscopy was performed in a closed-cycle helium cryostat (Monata Instruments CryoCore) with samples cooled to $<$5~\si{\kelvin}.
The sample was optically excited and the fluorescence collected using a long working distance infrared objective (Olympus LMPlan IR, 50$\times$/0.65 NA) with a home-built confocal microscopy setup.

\newpage

%\bibliographystyle{abbrv}
%\bibliography{references}
%%%%%%%%%%%%%%%%%%%%%%%%%%%%%%%%%%%%%%%%%%%%%%%%%%%%%%%%%%%%%%%%%%%%%
%% The same is true for Supporting Information, which should use the
%% suppinfo environment.
%%%%%%%%%%%%%%%%%%%%%%%%%%%%%%%%%%%%%%%%%%%%%%%%%%%%%%%%%%%%%%%%%%%%%

%%%%%%%%%%%%%%%%%%%%%%%%%%%%%%%%%%%%%%%%%%%%%%%%%%%%%%%%%%%%%%%%%%%%%
%% The "Acknowledgement" section can be given in all manuscript
%% classes.  This should be given within the "acknowledgement"
%% environment, which will make the correct section or running title.
%%%%%%%%%%%%%%%%%%%%%%%%%%%%%%%%%%%%%%%%%%%%%%%%%%%%%%%%%%%%%%%%%%%%%
\begin{acknowledgements}
R.U. acknowledges funding from the National Science Foundation (NSF) under the CAREER program (ECCS2339469) and a seed grant from the University of Iowa's Office of the Vice President for Research through the P3 Jumpstarting Tomorrow program.
N.D.B. thanks the Natural Sciences and Engineering Research Council of Canada (NSERC) for financially supporting this work under the Alliance International Catalyst Quantum grants program. 
The authors acknowledge support for carrying out the focused ion beam milling (FIB) and STEM work at the Canadian Centre for Electron Microscopy (CCEM), a national facility supported by McMaster University, the Ontario Research Fund (ORF), and the Canada Foundation for Innovation (CFI).
Pulsed laser deposition was conducted as part of a user project (CNMS2023-B-02072) at the Center for Nanophase Materials Sciences (CNMS), which is a US Department of Energy, Office of Science User Facility at Oak Ridge National Laboratory.

\end{acknowledgements}

%%%%%%%%%%%%%%%%%%%%%%%%%%%%%%%%%%%%%%%%%%%%%%%%%%%%%%%%%%%%%%%%%%%%%
%% The appropriate \bibliography command should be placed here.
%% Notice that the class file automatically sets \bibliographystyle
%% and also names the section correctly.
%%%%%%%%%%%%%%%%%%%%%%%%%%%%%%%%%%%%%%%%%%%%%%%%%%%%%%%%%%%%%%%%%%%%%
\bibliography{references_cleaned}

\end{document}